# Fabrication, phase formation and microstructure of $Ni_4Nb_2O_9$ ceramics by using two-stage sintering technique

**Orawan Khamman, Jiraporn Jainumpone, Anucha Watcharapasorn, Supon Ananta**

*Department of Physics and Materials Science, Faculty of Science,*

*Chiang Mai University, Chiang Mai 50200, Thailand*

The potential utilization of two-stage sintering for the production of highly dense and pure nickel diniobate ($Ni_4Nb_2O_9$) ceramics with low firing temperature was demonstrated. Effects of designed sintering conditions on phase formation, densification and microstructure of the ceramics were characterized by X-ray diffraction (XRD), Archimedes method and scanning electron microscopy (SEM), respectively. It has been found that minor phase of columbite $NiNb_2O_6$ tended to form together with the desired $Ni_4Nb_2O_9$ phase, depending on sintering conditions. The optimization of sintering conditions could lead to a single-phase $Ni_4Nb_2O_9$ ceramics with orthorhombic structure. The ceramics doubly sintered at 950/1250 °C for 4 h exhibited maximum density value of ~ 92%. Microstructures with denser angular grain-packing were generally found in both sets of the sintered $Ni_4Nb_2O_9$ ceramics. However, the grains were irregular in shape when the samples are sintered at 1050/1250 °C. Two-stage sintering was also found to enhance ferroelectric behavior of $Ni_4Nb_2O_9$ ceramic.

# I. INTRODUCTION

Nickel diniobate ($Ni_4Nb_2O_9$) which is one of the corundum-type $A_4Nb_2O_9$ (A = Mg, Co, Ni, Mn and Fe) ceramics exhibits orthorhombic symmetry with two different point groups (*i.e. Fd2d* and *Pbcn*). Its crystal structure can be represented as a network of edge- and corner-sharing $NiO_6$ and $NbO_6$ octahedra dimmers [1-3]. This material is a potential candidate for the development of dielectric ceramic resonators used at microwave frequencies especially for mobile communication and satellite systems [4-6]. Recently, Liou et al. prepared $Ni_4Nb_2O_9$ ceramics by the reaction-sintering process where a mixture of NiO and $Nb_2O_5$ precursors were directly sintered into $Ni_4Nb_2O_9$ ceramics by bypassing calcination step. However, they found that minor NiO and $NiNb_2O_6$ phases tended to form together with the $Ni_4Nb_2O_9$ phase, depending on sintering condition [7,8]. Thus, there has been a great deal of interest in the preparation of single-phase $Ni_4Nb_2O_9$ ceramic. In addition, the $Ni_4Nb_2O_9$ phase is very unstable, compared to the second phase that can easily be formed during preparation [7,8]. So far, these problems have been successfully resolved by applying a two-stage sintering technique in other similar systems where a potentially low-cost and simple ceramic fabrication to obtain highly dense and purely ceramics was demonstrated.

Therefore the two-stage sintering technique, instead of using a single firing at high temperature (up to 1350 °C) [8], is adopted. The aim of this study is to investigate the influence of these two ceramic processing methods (single- and two-stage sintering) on phase formation, microstructure and densification of $Ni_4Nb_2O_9$ ceramics.

# II. EXPERIMENTS AND DISCUSSION

The desired composition of $Ni_4Nb_2O_9$ ceramics was simply prepared by the solid-state reaction of appropriate amounts of commercially available nickel oxide (NiO) and niobium oxide ($Nb_2O_5$) (Aldrich, 99.9% purity). A ball-milling technique was employed for mixing the $Ni_4Nb_2O_9$ powders in isopropanal inert to the polypropylene jar. After drying at 120 °C for 24 h, the mixture was pressed into pellets and sintered in a closed alumina. $Ni_4Nb_2O_9$ ceramics was achieved by adding 2 wt% polyvinyl alcohol (PVA) binder, prior to pressing into pellets (10 mm in diameter and 1.0-1.5 mm in thickness) using a uniaxial die press at 100 MPa. After the binder was burned out at 500 °C for 1 h, green pellets were sintered in air at various dwell time (1-4 h) and heating/cooling rates of 10 and 20 °C/min. For the single-stage sintering samples, variation of the sintering temperature between 1200 and 1350 °C was carried out. Three sets of the first sintering temperature ($T_1$) were assigned for the two-stage sintering case: 950, 1050 and 1150 °C. Variation of the second sintering temperature ($T_2$) between 1200 °C and 1350 °C was carried out for each case. All ceramics were examined by room temperature X-ray diffraction (XRD; Siemens-D500 diffractometer), using Ni-filtered CuKα radiation, to identify the phases formed and optimum sintering condition for the production of single-phase $Ni_4Nb_2O_9$ ceramics. The relative amount of corundum and secondary phases was determined from XRD patterns of the samples by measuring the major characteristic peak intensities for the corundum (114) or $I_C$ and secondary phases or $I_S$. The following qualitative equation was used [9].

$$\text{Corundum phase}(\text{wt\%}) = \frac{I_c}{I_c + I_s} \times 100 \qquad (1)$$

The microstructure and grain sizes were directly imaged, using scanning electron microscopy (SEM; JEOL JSM-840A). The chemical compositions of the phase formed were elucidated by an energy-dispersive X-ray (EDX) analyzer with an ultra-thin window. EDX spectra were quantified with the virtual standard peaks supplied with the Oxford Instruments eXL software. Densities of the ceramics were measured by using the Archimedes principle. Ferroelectric hysteresis (P-E) loops were determined using a computer controlled modified Sawyer-Tower circuit at room temperature and 25 Hz [10]. Remanent polarization ($P_r$), saturation polarization ($P_{sat}$) and coercive field ($E_c$) were evaluated from the loops.

X-ray diffraction patterns from the single- and two-stage sintered $Ni_4Nb_2O_9$ ceramics are displayed in Fig. 1, indicating the formation of both corundum and impurity phases in each case. For the purpose of estimating the concentrations of the phase present, Eq. (1) has been applied to the XRD patterns, as given in Table 1. The results show that the desired $Ni_4Nb_2O_9$ phase has a corundum structure with an orthorhombic unit cell ($a$ = 871.9 pm, $b$ = 507.2 pm and $c$ = 1428.9 pm, space group *Pbcn* (no.60)), which could be matched with JCPDS file number 46-0525 [11] and was similar to other works [7]. In addition, the unwanted phase of $NiNb_2O_6$, which is in agreement with literature [7,8] has a columbite structure with an orthorhombic unit cell ($a$ = 1403 pm, $b$ = 568.7 pm and $c$ = 503.3 pm), consistent with JCPDS file number 76-2354 [12].

As the dwell time increased to 4 h, the intensity of the $NiNb_2O_6$ peaks was further reduced (Fig. 1). In this work, an evidence for the formation of $Ni_4Nb_2O_9$ phase, which coexisted with the $NiNb_2O_6$ phase, is found after firing at temperature ~ 1250-1300 °C, similar to those reported earlier [7,8]. From Fig. 1, it was obvious that the peaks of $NiNb_2O_6$ phase were completely eliminated after using heating/cooling rates of 20 °C/min. This finding was also consistent with our earlier work on $Ni_4Nb_2O_9$ powders system [7]. It was of interest to point out that for sintering temperature at 1350 °C, a single-phase of $Ni_4Nb_2O_9$ (yield of 100% within the limitation of the XRD technique) already formed. The observation that the dwell time and heating/cooling rates may also play an important role in obtaining a single-phase corundum ceramic, was also consistent with other similar systems [13]. XRD results suggested that the optimal sintering conditions for $Ni_4Nb_2O_9$ ceramics single- and two-stage sintered were 1350 °C and 950/1250 °C for 4 h with heating/cooling rates of 20 °C/min, respectively. Most importantly, it was evident that a two-stage sintering method could significantly lower the optimum sintering temperature for the formation of single-phase $Ni_4Nb_2O_9$ ceramics (Table 1).

To explore the roles of two-stage sintering on structural change and/or formation of intermediate phase ($NiNb_2O_6$) results obtained by XRD technique, microstructural features of single-and two-stage sintered $Ni_4Nb_2O_9$ samples were examined by using a combination of SEM and EDX technique, as shown in Fig. 2. In general, SEM micrographs (Fig. 2 (a) and (b)) of free surfaces revealed highly dense microstructures consisting of equiaxed grains and a few pores at triple points, in agreement with other works [8]. The grains fabricated by both sintering techniques were similar in shape. On contrary, the two-stage sintered $Ni_4Nb_2O_9$ sample had significantly smaller grain size and exhibited highly dense microstructure, consistent with other reports [14,15]. From Fig. 2 (b), the grain size of the two-stage sintered $Ni_4Nb_2O_9$ ceramics ranged 5-40 μm comparing to that observed in Fig. 2 (a), i.e. 10-70 μm. It could be confirmed by the SEM image (Fig. 2 (a)) that a loose formation of large grains occurred. The possible role of two-stage sintering for this observation is that the feasibility of densification without grain growth, which could be generated in the two-stage sintered ceramics, depended on the suppression of grain boundary movement while maintaining grain boundary diffusion activity [14,16]. The kinetics and the driving force for such grain growth characteristics were previously explained by Pakawanit et al. [14] and Chaisan et al. [15]. It was clearly seen that the sintering behavior of two different processing methods could give rise to different microstructures.

Moreover, it is seen that, when the $T_1$ temperature of this two-stage sintering method was increased to 1050 °C, $Ni_4Nb_2O_9$ grains are shaped abnormally (i.e. rod-like and equiaxed) as seen in Fig. 2 (c). The rod-like grains were generally ∼ 10-50 μm long and ∼ 5-7 μm wide. The presence of intermediate phase $NiNb_2O_6$ (also observed by XRD) with rod-like grains was confirmed by their chemical composition, as observed in the EDX spectrum marked as "(1)" and "(2)" (Fig. 2 (c) and Table 2). The existence of $NiNb_2O_6$ phase confirmed an expected problem of poor reactivity of NiO although the concentration was too low for detection by XRD, similar to our earlier work [7]. This finding demonstrated the advantage of a combination between SEM and EDX techniques, for the determination of the samples' microstructural features often undetected by the typical XRD diffraction which requires at least 5wt% of the component [17]. Even though exact mechanism of the microstructural development observed here could not yet be well established but it should be noted that the various features of microstructure in $Ni_4Nb_2O_9$ ceramics depended on the grain growth rate in the different planes. However, the sintering process and growth environment also played an important role in the formation of these microstructures. In this work, the occurrence of a typical abnormal-grain growth commonly found in the $Ni_4Nb_2O_9$ systems was observed. The density results could be correlated to the microstructure because high-density two-stage sintered ceramics exhibited high degrees of close packing of grains.

In addition, at higher firing temperatures in both sintering methods, 1350 and 1150/1350 °C, the grains seemed to partically melt as shown in Fig. 2 (d). It was found that the grains possessed more or less equiaxed shape and the grain boundaries could be clearly defined and deep trenches could be observed in certain areas. The melted grains may reduce porosity and resulting in higher density (as shown in Table 1); the higher the sintering temperature, the higher the density of the samples. The maximum relative density values of single-stage and two-stage sintered $Ni_4Nb_2O_9$ ceramics were about 92.7 and 92.6%, respectively (Table 1). For the first time, the polarization-electric field (P-E) hysteresis loops of $Ni_4Nb_2O_9$ ceramics prepared from different sintering conditions were shown in Fig. 3 and selected values were listed Table 3. Unlike in the single-stage sintered $Ni_4Nb_2O_9$ system where the non-hysteresis behavior was detected, all two-stage sintered samples exhibited slim-type hysteresis behavior which could partly be caused by effects of grain size [18]. Other factors may include the difference in polarizability in the corresponding structure and phase of ceramics prepared. Two-stage sintering ceramics in general showed large values of saturation polarization ($P_{sat}$) and remanent polarization ($P_r$) with smaller coercive field ($E_c$) than single-stage sintered sample. The presence of $NiNb_2O_6$, which may be a non-polarizable phase, could also have some effect on the observed ferroelectric behavior. This observation clearly clarifies effect of the co-existence of $Ni_4Nb_2O_9$ and $NiNb_2O_6$ phases, similar to the existence of the morphotropic phase boundary (MPB) in other ferroelectric systems, reported earlier [10].

## III. CONCLUSION

The two-stage sintering conditions were successfully employed to prepare $Ni_4Nb_2O_9$ ceramics. They were also found to have a pronounced effect on phase formation, microstructure, densification and ferroelectrics properties of $Ni_4Nb_2O_9$ ceramics. The secondary phase of $NiNb_2O_6$ disappeared when two-stage sintering was used instead of single-stage sintering when compared at the same maximum temperature. This work also demonstrated that it was possible to obtain rather dense and pure $Ni_4Nb_2O_9$ ceramics with better ferroelectrics properties at lower firing temperature by using two-stage sintering technique.

## ACKNOWLEDGEMENT

This work was supported by Advanced Metallic and Ceramics Materials Processing, Department of Physics and Materials Science, Faculty of Science, Chiang Mai University. Thank also for the Thailand Research Fund IRG5780013.

Table 1. Sintering behaviour of Ni$_4$Nb$_2$O$_9$ ceramics at various sintering conditions

| $T_1$ (°C), dwell time (h), rate (°C/min) | $T_2$ (°C), 4 h, 20 °C/min | Corundum Ni$_4$Nb$_2$O$_9$ phase (%) | Relative density (%) |
| --- | --- | --- | --- |
| 950, 4, 20 | 1200 | 85.0 | 88.5 |
| 950, 4, 20 | 1250 | 100.0 | 92.2 |
| 1050, 4, 20 | 1200 | 81.9 | 90.0 |
| 1050, 4, 20 | 1250 | 100.0 | 91.1 |
| 1150, 4, 20 | 1200 | 77.9 | 90.8 |
| 1150, 4, 20 | 1250 | 100.0 | 92.0 |
| 1150, 4, 20 | 1350 | 100.0 | 92.6 |
| 1200, 4, 20 | - | 83.2 | 90.0 |
| 1250, 4, 20 | - | 85.4 | 91.4 |
| 1300, 1, 10 | - | 75.5 | 85.0 |
| 1300, 4, 10 | - | 80.0 | 91.0 |
| 1300, 4, 20 | - | 90.0 | 91.5 |
| 1350, 4, 20 | - | 100.0 | 92.7 |

Table 2. Chemical compositions of Ni$_4$Nb$_2$O$_9$ ceramics sintered at 1050/1250 °C for 4 h with heating/cooling rates 20 °C/min (Fig 2 (c))

| EDX positions | Composition (at %) | | Possible phase |
| --- | --- | --- | --- |
| | Ni (K) | Nb (L) | |
| (1) | 22.97 | 10.55 | Ni$_4$Nb$_2$O$_9$ |
| (2) | 14.28 | 27.29 | NiNb$_2$O$_6$ |

Table 3. Ferroelectric properties of $Ni_4Nb_2O_9$ ceramics sintered at various conditions

| sintering temperature (°C) | | ferroelectric properties (at 25 °C, 25 Hz) | | |
| --- | --- | --- | --- | --- |
| $T_1$ | $T_2$ | $P_s$ (μC/cm$^2$) | $P_r$ (μC/cm$^2$) | $E_c$ (kV/cm) |
| 1200 | - | 1.16 | 0.72 | 17.80 |
| 1250 | - | 1.62 | 0.92 | 18.34 |
| 1350 | - | 1.81 | 0.97 | 17.44 |
| 950 | 1250 | 38.78 | 4.37 | 2.63 |
| 1050 | 1250 | 76.20 | 8.95 | 2.43 |
| 1150 | 1250 | 52.15 | 7.18 | 2.53 |

**Figure Captions.**

Fig. 1. XRD patterns of $Ni_4Nb_2O_9$ ceramics singly and doubly sintered at various conditions.

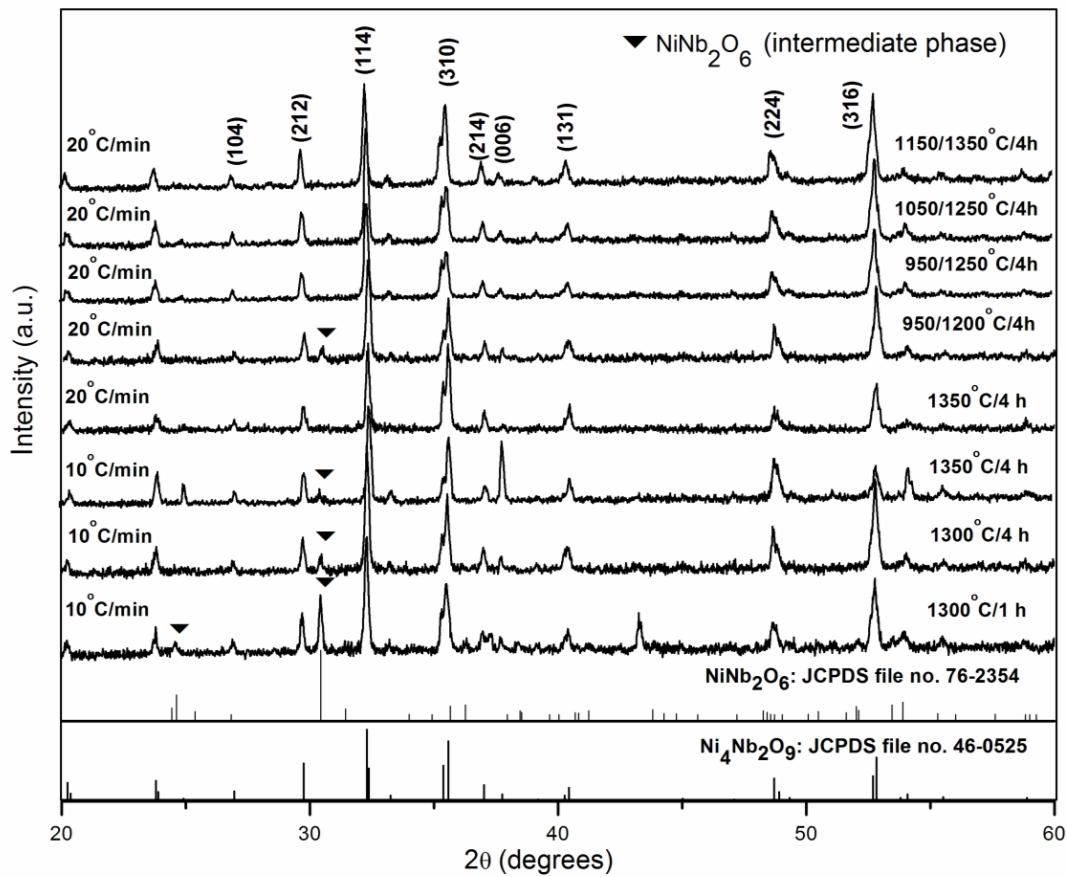

Fig. 2. SEM micrographs of Ni$_4$Nb$_2$O$_9$ ceramics sintered at (a) 1250 °C, (b) 950/1250 °C, (c) 1050/1250 °C and (d) 1150/1350 °C for 4 h with heating/cooling rates 20 °C/min.

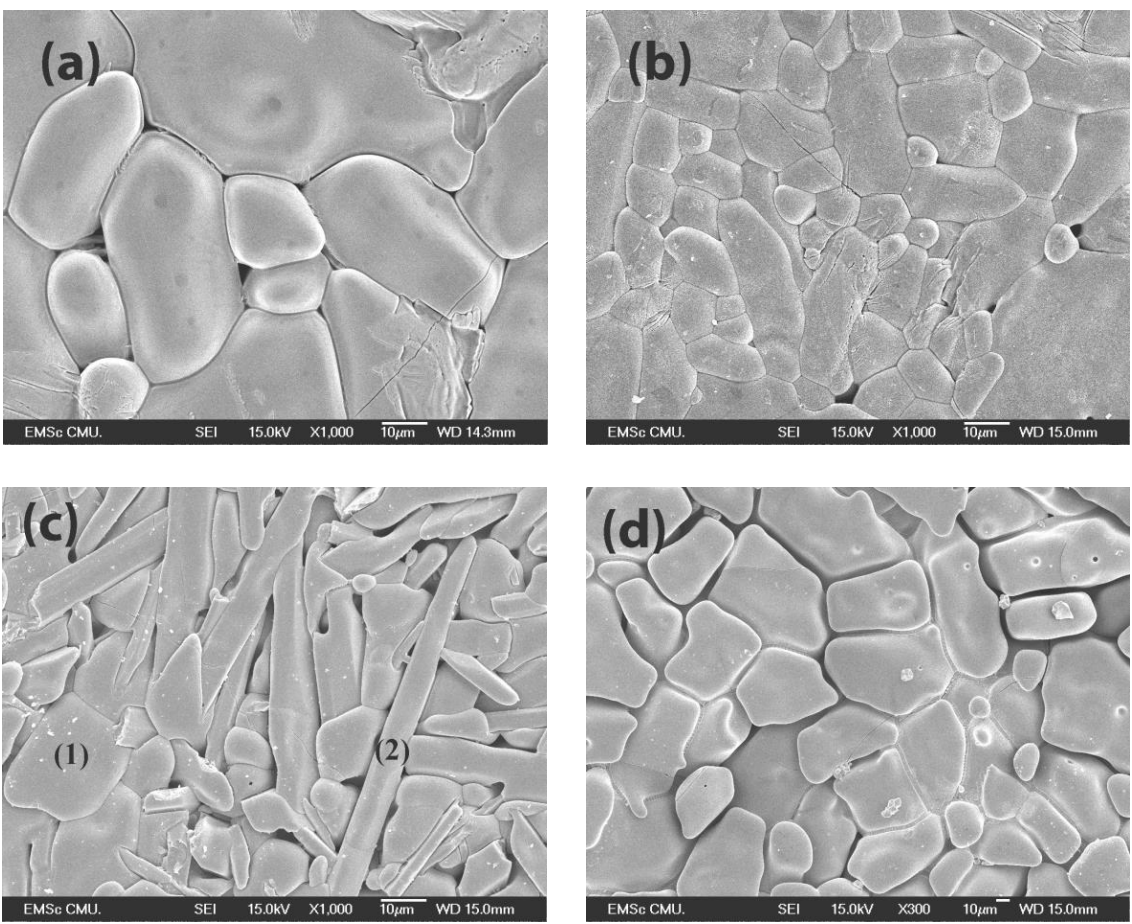

Fig. 3. P-E hysteresis loops of sintered $Ni_4Nb_2O_9$ ceramics

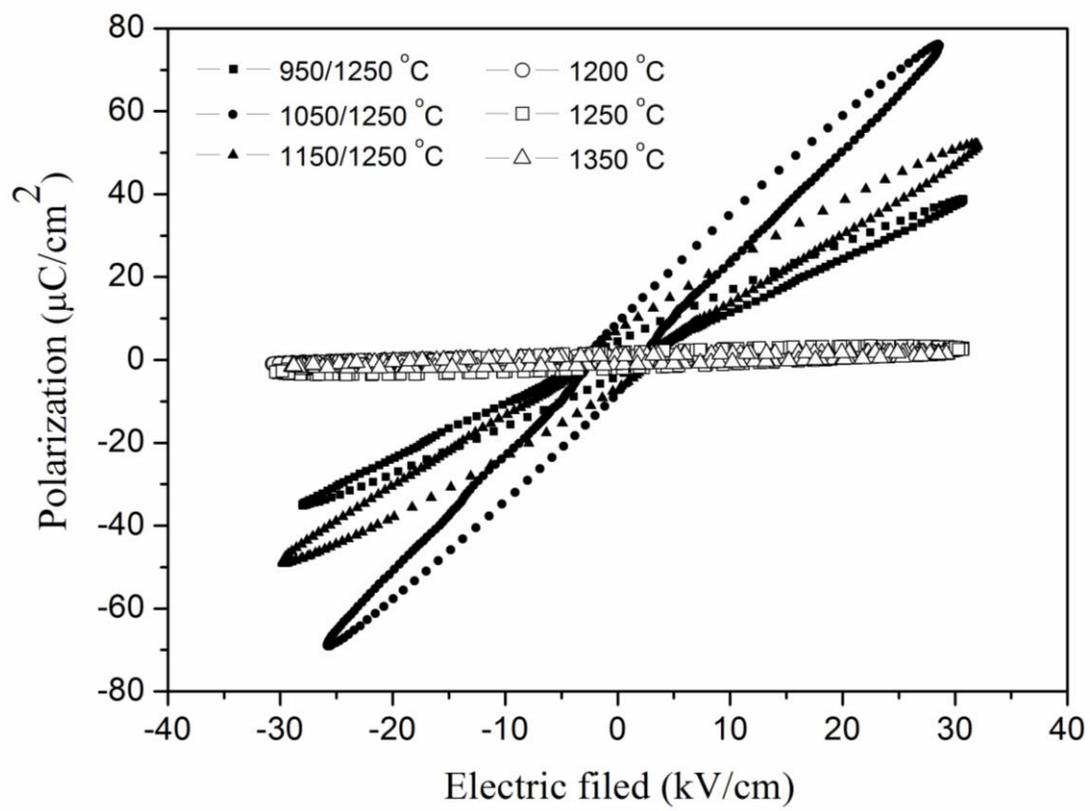